\documentclass[conference]{IEEEtran}%
\IEEEoverridecommandlockouts
\usepackage{amsfonts, amssymb}
\usepackage{amsmath}
\usepackage{cite}
\usepackage[pdftex]{graphicx}
\usepackage[T1]{fontenc}
\usepackage{balance}%
\usepackage{amsfonts}%
\usepackage{amssymb}
\usepackage{xcolor}
\usepackage{lipsum}
\usepackage{mathtools}
\usepackage{cuted}
\usepackage{amsthm}
\usepackage{subcaption}
\usepackage{float}
\usepackage[font=small, labelsep=none]{caption}
\usepackage{multicol}
\usepackage{graphicx}
\usepackage{flushend}

\usepackage{algorithm}
\usepackage{algpseudocode}
\usepackage{amsthm}  
\usepackage{amsthm}
\newtheorem{remark}{Remark}
\newtheorem{assumption}{Assumption}
\providecommand{\U}[1]{\protect\rule{.1in}{.1in}}
\begin{document}

\title{Analog Over-the-Air Federated Learning with Interference-Based Energy Harvesting}

\author{
    \normalsize
    Ahmad~Massud~Tota~Khel\textsuperscript{1}, 
    Aissa~Ikhlef\textsuperscript{1}, 
    Zhiguo~Ding\textsuperscript{2}, 
    Hongjian~Sun\textsuperscript{1}\\
    \small
    \textsuperscript{1}Department of Engineering, Durham University, Durham, U.K.\\
    \textsuperscript{2}Department of Electrical and Electronic Engineering, The University of Manchester, Manchester, U.K.\\
    \textsuperscript{1}\{ahmad.m.tota-khel, aissa.ikhlef, hongjian.sun\}@durham.ac.uk, 
    \textsuperscript{2}zhiguo.ding@manchester.ac.uk
    \thanks{This work was supported by the CHEDDAR: Communications Hub for Empowering Distributed ClouD Computing Applications and Research funded by the UK EPSRC under grant numbers EP/Y037421/1 and EP/X040518/1.}
}

\maketitle
\setlength{\columnsep}{0.2in}
\begin{abstract}
We consider analog over-the-air federated learning, where devices harvest energy from in-band and out-band radio frequency signals, with the former also causing co-channel interference (CCI). To mitigate the aggregation error, we propose an effective denoising policy that does not require channel state information (CSI). We also propose an adaptive scheduling algorithm that dynamically adjusts the number of local training epochs based on available energy, enhancing device participation and learning performance while reducing energy consumption. Simulation results and convergence analysis confirm the robust performance of the algorithm compared to conventional methods. It is shown that the performance of the proposed denoising method is comparable to that of conventional CSI-based methods. It is observed that high-power CCI severely degrades the learning performance, which can be mitigated by increasing the number of active devices, achievable via the adaptive algorithm. 
\end{abstract} 

\markboth{IEEE }{Massud \MakeLowercase{\textit{et al.}}: IEEE }

\IEEEpeerreviewmaketitle
\vspace{-8.5pt}\section{Introduction}\vspace{-2pt}
Future sixth-generation (6G) wireless networks are envisioned to support large-scale intelligent connectivity, where a massive number of Internet-of-Things (IoT) devices collaborate to enable real-time and privacy-aware learning\cite{9562487,10109680}. As such, federated learning (FL) emerges as a key enabler by allowing distributed IoT devices to collaboratively train local models and share model updates instead of raw data\cite{9264742}. However, traditional digital FL systems in dense IoT networks—relying on separate uplink transmissions from each device to a parameter server (PS)—incur significant communication overhead and high energy consumption\cite{10696917,9834342}. Thus, analog over-the-air (OTA) aggregation has emerged as a promising solution to reduce this overhead by enabling simultaneous transmissions from multiple devices over the same frequency band, leveraging the superposition property of the wireless channel\cite{9562487,10109680}. Still, communication overhead is only part of the challenge, as powering a massive number of IoT devices using batteries or wired supplies is costly, hard to maintain, and environmentally unsustainable. As a more sustainable alternative, energy harvesting (EH) enables IoT devices to operate by extracting energy from ambient sources (e.g., radio frequency (RF) signals), thereby reducing maintenance needs and harmful environmental impact\cite{10000620,9453711,11029479}.

Numerous studies have explored OTA FL systems \cite{10109680,10000620,9562487,9834342,9672092,9158563,10767214,9798757,10696917}, but have ignored the impact of co-channel interference (CCI) on the aggregation and convergence, even in those involving EH-based devices \cite{10000620,9453711}, thereby limiting their practicality, especially in dense IoT environments. Moreover, to mitigate the aggregation error, most of them have applied either the power-hungry channel inversion technique to adjust the transmit power at devices, or denoising factors at the PS, which require channel state information (CSI). However, considering the energy constraints of IoT devices and the complexity of acquiring CSI, these methods limit the efficiency and scalability of FL systems. In addition, most of these works rely on a fixed number of local training epochs, which limits their flexibility in energy-constrained environments \cite{10109680,9264742,10000620,9562487}. Although in \cite{9453711} the number of active devices is optimized based on the harvested energy, each device still performs a fixed number of epochs on full datasets. Similarly, \cite{9834342} adapts the number of epochs based on power constraints but assumes a fixed number of active devices without EH capabilities.

Motivated by these considerations, we consider an analog OTA FL system where devices harvest energy from RF signals of coexisting communication nodes across various frequency bands: (i) inband signals, overlapping with the system’s operating frequencies and causing CCI at the PS; and (ii) outband signals, operating on separate bands without causing CCI. To avoid the power-hungry channel inversion at the devices, we propose a CSI-free denoising policy—variance-based denoising—applied by the PS, accounting for the effects of fading, CCI, and additive white Gaussian noise (AWGN). Moreover, to improve energy efficiency and convergence, we propose an adaptive algorithm that dynamically adjusts the number of epochs per device based on the available energy, enabling fractional dataset processing when full training is infeasible and increasing device participation. To examine the effects of the proposed algorithm and denoising policy on learning performance, we present a theoretical convergence analysis.

Simulation results demonstrate that the proposed denoising policy achieves performance comparable to conventional mean squared error (MSE)-based and fading-based policies, which require CSI. The results also confirm the robustness of the proposed adaptive algorithm in comparison to conventional non-adaptive local training methods. It is demonstrated that the proposed algorithm not only accelerates the convergence but also reduces the overall energy consumption. It is also observed that the learning performance is significantly degraded by high-power CCI, which can be mitigated by increasing the number of active devices, a mitigation that can be achieved through the proposed adaptive algorithm.

The rest of the paper is organized as follows. The system model, denoising policies and adaptive algorithm, convergence analysis, simulation results, and conclusion of the paper are presented in Sections \ref{sec2}, \ref{sec3}, \ref{sec4}, \ref{sec5}, and \ref{sec6}, respectively.  
\section{System Model}\label{sec2}
We consider a wireless FL system that employs analog OTA aggregation to collaboratively train a global machine learning model. The system comprises $M$ distributed devices that communicate with a PS over $T$ communication rounds. At the start of the $t$-th round, where $t \in \{1, \dots, T\}$, the PS broadcasts the global model parameter vector $\mathbf{w}_t \in \mathbb{R}^d$—where $d$ is the number of trainable parameters—to all devices in the downlink. The PS is assumed to use dedicated energy sources, providing reliable power to support the widely adopted assumption of error-free downlink parameter transmission \cite{9834342,10109680}. The objective of the global learning process at round $t$ is to minimize the global loss function, defined as
\begin{align}
    F(\mathbf{w}_t) = \frac{1}{\sum_{m=1}^{M} |\mathcal{D}_m|} \sum_{m=1}^{M} |\mathcal{D}_m| F_m(\mathbf{w}_t), \label{Glob1}
\end{align}
where $F_m(\mathbf{w}_t)$ is the local loss function, defined as
\begin{align}
    F_m(\mathbf{w}_t) = \frac{1}{|\mathcal{D}_m|} \sum_{u \in \mathcal{D}_m} f(\mathbf{w}_t, u),
\end{align}
where $\mathcal{D}_m$ is the local dataset of device $m$, and $f(\mathbf{w}_t, u)$ is the loss function for a data sample $u$ with respect to $\mathbf{w}_t$.

Each device runs local stochastic gradient descent (SGD) for $\tau_m$ epochs to minimize $F_m(\mathbf{w}_t)$. Letting $\mathbf{w}_{m,t}^{0} \triangleq \mathbf{w}_t$, the local update rule at the $j$-th epoch of round $t$ is written as\cite{10109680}
\begin{align}
    \mathbf{w}_{m,t}^{(j+1)} = \mathbf{w}_{m,t}^{j} - \eta \nabla F_m(\mathbf{w}_{m,t}^{j}), \ j = 0, \dots, \tau_m - 1.
\end{align}

After local training in communication round $t$, the devices share their model differences with the PS, defined as \cite{10109680,10000620}
\begin{align}
    \Delta \mathbf{w}_{m,t} = \mathbf{w}_t - \mathbf{w}_{m,t}^{\tau_m}.\label{modediff}
\end{align}

We assume that the devices with broadband EH circuits operate by extracting energy from RF signals of \( I \) inband and \( K \) outband coexisting communication nodes. Let $\mathbf{u}_{i,t} \in \mathbb{C}^d$ and $\mathbf{v}_{k,t} \in \mathbb{C}^d$ denote zero-mean, unit-power complex Gaussian signals from inband and outband sources, respectively. For device \( m \), the fading channel and distance to the \( i \)-th inband node are \( h_{m,i,t}^{\mathrm{in}}\! \sim\! \mathcal{CN}(0,1) \) and \( d_{m,i}^{\mathrm{in}} \); those to the \( k \)-th outband node are \( h_{m,k,t}^{\mathrm{out}} \!\sim\! \mathcal{CN}(0,1) \) and \( d_{m,k}^{\mathrm{out}} \). The harvested energy by the $m$-th device in communication round $t$ is written as \cite{11029479}
\begin{align}
   E_{m,t} =T^{\text{h}} \delta_m \Big(\sum_{i=1}^I \mathcal{L}^{\mathrm{in}}_i \big|h_{m,i,t}^{\mathrm{in}}\big|^2 +\sum_{k=1}^K \mathcal{L}^{\mathrm{out}}_k  \big|h_{m,k,t}^{\mathrm{out}}\big|^2\Big),\label{HarvestedE}
\end{align}
where $T^{\text{h}}$ is the duration of round $t$, $\delta_m \in (0,1]$ is the energy conversion efficiency, \(\mathcal{L}^{\mathrm{in}}_i=P^{\mathrm{in}}_i (d_{m,i}^{\mathrm{in}})^{-\xi}\), \(\mathcal{L}^{\mathrm{out}}_k=P^{\mathrm{out}}_k (d_{m,k}^{\mathrm{out}})^{-\xi}\), $\xi$ is the path-loss exponent, and $P^{\mathrm{in}}_i$ and $P^{\mathrm{out}}_k$ are the transmit powers of the inband and outband nodes.

We can quantify the total energy consumption of the $m$-th device in communication round $t$ as \cite{9264742} 
\begin{align}
    E_{m,t}^{\text{Cons}} = E_{m,t}^{\text{up}} + \tau_m E_m^{\text{comp}},
\end{align}
where $E_{m,t}^{\text{up}}$ denotes the energy allocated for the uplink transmission, and $E_m^{\text{comp}}=\kappa C_m |\mathcal{D}_m| f_m^2$ is the energy consumed per epoch for local computation. Here, $\kappa$ is the effective switched capacitance, $C_m$ is the number of CPU cycles per sample, and $f_m$ is the processor frequency \cite{9264742}. 

We assume that each device is equipped with a battery of finite capacity $B_{\max}$. At the start of the $t$-th communication round, the $m$-th device has a battery energy level denoted by $B_{m,t}$, which consists solely of harvested energy stored from previous communication rounds. The device uses this stored energy to perform local computation and transmit its model difference during communication round $t$. Meanwhile, it continuously harvests energy throughout the communication round, $E_{m,t}$, which is added to the battery at the end of the round and becomes available for use starting from round $(t{+}1)$. Accordingly, the battery energy level is updated as \cite{9453711}
\vspace{-8pt}\begin{align}
    B_{m,(t+1)} \!=\! \min\left\{B_{\max}, B_{m,t} - E_{m,t}^{\text{Cons}} + E_{m,t}\right\}.\label{battlevel}
\end{align}

A device is considered eligible to participate in communication round $t$ if its available energy satisfies $B_{m,t} \geq E_{m,t}^{\text{Cons}}$; otherwise, it remains idle and stores the harvested energy in its battery. We define $a_{m,t} \in \{0,1\}$ as a binary variable indicating the activity of device $m$ in round $t$, which is expressed as \cite{10000620}
\vspace{-14pt}\begin{align}
    a_{m,t} = 
    \begin{cases}
    1, & \text{if } B_{m,t} \geq E_{m,t}^{\text{Cons}}, \\
    0, & \text{otherwise}.
    \end{cases}
\end{align}

Let $\mathcal{A}_t \subseteq \{1, \dots, M\}$ be the set of active devices at communication round $t$, with cardinality $|\mathcal{A}_t| = \sum_{m=1}^{M} a_{m,t}=N_{t} \leq M$. Thus, the global loss function given in (\ref{Glob1}) is rewritten as
\vspace{-2.5pt}\begin{align}
    F(\mathbf{w}_t) = \frac{1}{\sum_{m \in \mathcal{A}_{t}} |\mathcal{D}_m|} \sum_{m \in \mathcal{A}_{t}} |\mathcal{D}_m| F_m(\mathbf{w}_t).
\end{align}

Following the OTA strategy, all active devices transmit their local model differences, $\Delta \mathbf{w}_{m,t}$, simultaneously over the same uplink band. To enable coherent aggregation, each device applies phase alignment and embeds the model update into the transmit signal, which is written as \cite{10109680,9158563} 
\vspace{-2.5pt}\begin{align}
    \mathbf{x}_{m,t}  = \frac{h_{m,t}^{*}}{|h_{m,t}|}  \sqrt{P_{m,t}^{\text{up}}} \Delta \mathbf{w}_{m,t}, \  m\in \mathcal{A}_t  ,\label{txsigg}
\end{align}
where $h_{m,t}\! \sim\!\mathcal{CN}(0,1)$ is the uplink fading channel between device $m$ and the PS, $h_{m,t}^{*}$ is its conjugate, $P_{m,t}^{\text{up}}\!=\!E_{m,t}^{\text{up}}/T_m^{\text{up}}$ is the transmit power, and $T_m^{\text{up}}$ is the uplink duration.

The received signal at the PS in communication round $t$, affected by inband CCI signals, is written as
\vspace{-2.5pt}\begin{align}
    \mathbf{y}_{t} &=\sum_{m \in \mathcal{A}_{t}} \sqrt{P_{m,t}^{\text{up}} d_m^{-\xi}} |h_{m,t}| \Delta \mathbf{w}_{m,t}\nonumber\\
    &+ \sum_{i=1}^{I} \sqrt{P^{\mathrm{in}}_i (d_{i}^{\mathrm{in}})^{-\xi}} g_{i,t} \mathbf{u}_{i,t}
    + \mathbf{z}_t,\label{recsig}
\end{align}
where $d_m$ is the distance from device $m$ to the PS, $g_{i,t} \!\sim\! \mathcal{CN}(0,1)$ and $d_{i}^{\mathrm{in}}$ are the fading channel and distance from the $i$-th CCI to the PS, and $\mathbf{z}_t \!\sim\! \mathcal{CN}(0, N_0 \mathbf{I}_d)$ is the AWGN.

\section{Denoising Policy and Adaptive FL Algorithm}\label{sec3}
\subsection{Aggregated Update and Denoising Policy}
We assume that the PS employs a denoising policy to mitigate the aggregation error \cite{10767214}. By using this policy, energy-constrained devices avoid the conventional channel inversion technique, which not only requires more energy but also fails to mitigate CCIs and AWGN, as it is a pre-transmission process \cite{10109680,10767214,9562487}. Thus, the aggregated update with a denoising factor, $\alpha_{t}$, is written as \cite{10109680} 
\begin{align}
    \hat{\mathbf{s}}_{t}=\frac{\mathbf{y}_{t}}{\alpha_{t} N_{t}}.\label{aggden}
\end{align}

It is to note that the ideal aggregated update in communication round $t$ is expressed as \cite{9562487}
\begin{align}
    \mathbf{s}_{t} = \frac{1}{N_{t}} \sum_{m \in \mathcal{A}_{t}} \Delta \mathbf{w}_{m,t}.\label{idagg1}
\end{align}

Therefore, using (\ref{recsig}), (\ref{aggden}), and (\ref{idagg1}), the aggregation error can be written as \cite{9562487}
\begin{align}
    \hat{\mathbf{s}}_{t} \!-\! \mathbf{s}_{t} =& \frac{1}{ N_t} \!\sum_{m \in \mathcal{A}_{t}} \Big(\frac{1}{\alpha_{t}}\sqrt{P_{m,t}^{\text{up}}d_{m}^{-\xi}} |h_m (t)| \!-\! 1\Big) \Delta \mathbf{w}_{m,t} \nonumber\\
    &\!+\! \frac{1}{\alpha_{t}N_t} \Bigg(\!\sum_{i=1}^{I}\!\sqrt{P^{\mathrm{in}}_i (d_{i}^{\mathrm{in}})^{-\xi}} g_{i,t} \mathbf{u}_{i,t}
    \!+\! \mathbf{z}_t\Bigg). \label{error_term_exp}
\end{align}

We consider three different denoising methods: (i) fading-based, (ii) MSE-based, and (iii) variance-based. 
\subsubsection{Fading-Based Denoising} We assume that the PS has the CSI of active devices and employs a denoising factor to compensate for the effects of fading and path-loss \cite{9798757}. As a result, using (\ref{error_term_exp}), similar to \cite[Proposition 2]{9798757}, the optimal denoising factor for such a case can be expressed as
\begin{align}
    \alpha_{t} = \frac{1 }{N_{t} }\sum_{m \in \mathcal{A}_{t}} \sqrt{P_{m,t}^{\text{up}}d_{m}^{-\xi}} |h_{m,t}|.\label{optden}
\end{align}
\subsubsection{MSE-Based Denoising} To improve the model convergence and aggregation error, most existing works derive a denoising factor that minimizes the MSE, $||\hat{\mathbf{s}}_{t} - \mathbf{s}_{t}||^{2}$ \cite{10109680,9158563,10767214,9562487,9798757}. This method requires CSI for both active devices and CCIs, where the MSE is written as\cite[Eq.~(8)]{9158563}
\begin{align}
    \text{MSE}_{t} \!=\! \frac{d}{ N_t^{2}} \!\sum_{m \in \mathcal{A}_{t}} \!\Big(\frac{1}{\alpha_{t}}\sqrt{P_{m,t}^{\text{up}}}d_{m}^{-\frac{\xi}{2}} |h_{m,t}| \!-\! 1\Big)^{2} \!+\! \frac{d\varphi_{t}}{\alpha_{t}^{2} N_t^{2}} , \label{MSEC}
\end{align}
where \(\varphi_{t}=\sum_{i=1}^{I} P^{\mathrm{in}}_i (d_{i}^{\mathrm{in}})^{-\xi} |g_{i,t}|^{2}+N_{0}\).

Using \cite[Appendix B]{9158563}, the optimal $\alpha_{t}$ is obtained as
\begin{align}
    \alpha_{t} = \frac{\sum_{m \in \mathcal{A}_{t}} P_{m,t}^{\text{up}}d_m^{-\xi}|h_{m,t}|^{2}+\varphi_{t}}{\sum_{m \in \mathcal{A}_{t}} \sqrt{P_{m,t}^{\text{up}}}d_{m}^{-\frac{\xi}{2}} |h_{m,t}|}.\label{MSEDEN}
\end{align}

\subsubsection{Variance-Based Denoising}
Obtaining accurate CSI, especially in large-scale wireless FL systems with CCI, poses significant challenges due to user mobility, feedback overhead, and privacy constraints \cite{9158563}. To address this, we propose a variance-based denoising method that eliminates the need for CSI by normalizing the received aggregated signal using its standard deviation, which inherently captures the combined effects of the desired signal, CCI, and AWGN. Since the received signal $\mathbf{y}_{t} \in \mathbb{R}^d$ is a random vector representing the superposition of zero-mean independent signals, using its per-dimension variance, $\mathbb{V}[\mathbf{y}_{t}] = \frac{1}{d} \mathbb{E}[\|\mathbf{y}_{t}\|^2]$, (\ref{recsig}), and (\ref{aggden}), the variance-based denoising factor can be expressed as
\begin{align}
    \alpha_{t} = \sqrt{\mathbb{V}\left[\frac{\mathbf{y}_{t}}{N_{t}}\right]} = \frac{1}{N_{t}} \sqrt{ \sum_{m \in \mathcal{A}_{t}} P_{m,t}^{\text{up}} d_m^{-\xi} + \varphi_{t} } \ .\label{VARDEN}
\end{align}

\subsection{Energy-Efficient Adaptive OTA FL Algorithm}
We provide Algorithm \ref{adalg} that dynamically adjusts the number of epochs on each device based on the available energy. When the available energy is insufficient for full dataset training, the algorithm supports the use of fractional datasets. As a result, this adaptive approach increases device participation per communication round, speeds up global model convergence, and reduces total energy consumption.
\begin{algorithm}[H]
\caption{Adaptive OTA FL Algorithm}
\label{adalg}
\begin{algorithmic}[1]
\State \textbf{Input:} $T$, $M$, $\eta$, $\mathcal{D}_m\forall m\in\{1,\dots,M\}$, $E_{m,t}^{\text{up}}$ 
\State \textbf{Initialize:} Initial global model parameters $\mathbf{w}_1$, initial battery levels $B_{m,1}$, and calculate $E_m^{\text{comp}} = \kappa C_m |\mathcal{D}_m| f_m^2$
\For{each communication round $t = 1$ to $T$}
    \State PS broadcasts model $\mathbf{w}_{t}$ to all devices

    \For{each device $m = 1$ to $M$ \textbf{in parallel}}

        \State Calculate the harvested energy, $E_{m,t}$, using (\ref{HarvestedE})
        
        \If{$B_{m,t} \geq E_{m,t}^{\text{up}} + E_m^{\text{comp}}$}
            \State Allocate full epochs: $\tau_m \!\gets\! \left\lfloor \dfrac{B_{m,t} - E_{m,t}^{\text{up}}}{E_m^{\text{comp}}} \right\rfloor$
            \State Use full dataset $\mathcal{D}_m$,  and $a_{m,t} \gets 1$
            \State Update the battery energy level using (\ref{battlevel})

        \ElsIf{$B_{m,t} > E_{m,t}^{\text{up}}$}
            \State $r_{m,t} \gets \dfrac{B_{m,t} - E_{m,t}^{\text{up}}}{E_m^{\text{comp}}}$
            \State Select subset $\mathcal{D}_m^f$ of size $\lfloor r_{m,t}  |\mathcal{D}_m| \rfloor$
            \State Set $\tau_m \gets 1$, $a_{m,t} \gets 1$
            \State Update the battery energy level: $B_{m,(t+1)} \gets \min\left\{ B_{\max}, B_{m,t} - E_{m,t}^{\text{up}} - r_{m,t} E_m^{\text{comp}}+E_{m,t} \right\}$

        \Else
            \State Update the battery energy level: $B_{m,(t+1)} \gets \min\left\{ B_{\max}, B_{m,t}+E_{m,t} \right\}$, and $a_{m,t} \gets 0$
        \EndIf

        \If{$a_{m,t} = 1$}
            \State Train for $\tau_m$ epochs on assigned data
            \State Compute model difference: $\Delta \mathbf{w}_{m,t} $ using (\ref{modediff})
        \EndIf

    \EndFor

    \State Devices simultaneously transmit $\mathbf{x}_{m,t}$ given in (\ref{txsigg}) 
    \State PS aggregates and denoises $\mathbf{y}_{t}$, as given in (\ref{aggden})
    \State PS updates the global model: $\mathbf{w}_{t+1} \gets \mathbf{w}_{t} - \hat{\mathbf{s}}_{t}$

\EndFor

\State \Return Final model $\mathbf{w}_{T+1}$

\end{algorithmic}
\end{algorithm}

\section{Convergence Analysis}\label{sec4}
We analyze the convergence of the FL system by bounding the average expected squared gradient norm, indicating convergence to a stationary point. The analysis relies on standard assumptions commonly used in prior works \cite{10109680,9158563,9562487,10696917}.
\begin{assumption}
Each local loss function, $F_m(\mathbf{w}_t)$, is $L$-smooth, i.e., for any $\{\mathbf{w}_t, \mathbf{w}_t'\} \in \mathbb{R}^d$, the following holds:
\begin{align}
F_m(\mathbf{w}_t) \leq F_m(\mathbf{w}_t') \!+\! \langle \nabla F_m(\mathbf{w}_t'), \mathbf{w}_t \!-\! \mathbf{w}_t' \rangle \!+\! \frac{L}{2} \|\mathbf{w}_t \!-\! \mathbf{w}_t'\|^2.
\end{align}
\end{assumption}
\begin{assumption} Each local gradient is an unbiased estimator of the global gradient, defined as
\begin{align}
\mathbb{E}[\nabla F_m(\mathbf{w}_t)] = \nabla F(\mathbf{w}_t), \ \forall m,\mathbf{w}_t.
\end{align}
\end{assumption}
\begin{assumption}
The local gradients' norm is bounded as
\begin{align}
\|\nabla F_m(\mathbf{w}_t)\|^2 \leq G^2, \quad \forall \mathbf{w}_t, m,
\end{align}
where $G^2$ is a non-negative constant upper bound.  
\end{assumption}

\begin{assumption}
The denoised aggregated model difference satisfies the following second-moment error bound:
\begin{align}
\mathbb{E}\left[\|\hat{\mathbf{s}}_{t} - \mathbf{s}_{t}\|^2\right] \leq \zeta_t^2,
\end{align}
where \( \zeta_t^2 \geq 0 \) is a constant upper bound on the aggregation noise power in communication round \( t \).
\end{assumption}

Using Assumption 1 and \cite[Lemma 1]{10109680}, the global loss function, $F(\mathbf{w}_t)$, as the average of local loss functions, inherits the $L$-smoothness property as
\begin{align}
F(\mathbf{w}_{t+1}) &\!\leq\! F(\mathbf{w}_t) \!+\! \langle \nabla\! F(\mathbf{w}_t), \mathbf{w}_{t+1} \!-\! \mathbf{w}_t \rangle \!+\! \frac{L}{2} \|\mathbf{w}_{t+1} \!-\! \mathbf{w}_t\|^2 \!.\!\label{globLS1}
\end{align}

We then use the global model update rule, \(\mathbf{w}_{t+1}=\mathbf{w}_{t} - \hat{\mathbf{s}}_{t}\), and take the expectation of (\ref{globLS1}), which yields
\begin{align}
\mathbb{E}[F(\mathbf{w}_{t+1})]
&\leq \mathbb{E}[F(\mathbf{w}_t)] \!-\! \mathbb{E}[\langle \nabla\! F(\mathbf{w}_t), \hat{\mathbf{s}}_{t} \rangle] \!+\! \frac{L}{2} \mathbb{E}[\|\hat{\mathbf{s}}_{t}\|^2].\label{expmain}
\end{align}

Following the SGD rule given in (\ref{modediff}), and using Assumption 2, the ideal aggregated update given in (\ref{idagg1}) is rewritten as
\begin{align}
\mathbf{s}_{t} \!=\! \frac{1}{N_t} \!\sum_{m \in \mathcal{A}_{t}} \!\Delta \mathbf{w}_{m,t} \!=\! \frac{\eta}{N_t} \!\sum_{m \in \mathcal{A}_{t}} \!\sum_{j=0}^{\tau_m - 1} \nabla\! F_m(\mathbf{w}_{m,t}^{j}).\label{newideal}
\end{align}

Using (\ref{newideal}) and the triangle inequality, it can be written that
\begin{align}
    \|\mathbf{s}_{t}\|^2 &= \Big\| \frac{\eta}{N_t} \sum_{m \in \mathcal{A}_{t}} \sum_{j=0}^{\tau_m - 1} \nabla F_m(\mathbf{w}_{m,t}^j) \Big\|^2 \nonumber\\
&\leq \Big( \frac{\eta}{N_t} \sum_{m \in \mathcal{A}_{t}} \sum_{j=0}^{\tau_m - 1} \left\| \nabla F_m(\mathbf{w}_{m,t}^j) \right\| \Big)^2.
\end{align}

By applying Assumption 3 and taking the expectation, the final bound on the expected squared norm is obtained as
\begin{align}
  \mathbb{E}[\|\mathbf{s}_{t}\|^2] \leq \Big( \frac{\eta}{N_t} \sum_{m \in \mathcal{A}_{t}} \tau_m G \Big)^2=  \eta^2 \bar{\tau}_{t}^2 G^2,  \label{exps2}
\end{align}
where \( \bar{\tau}_{t} \!=\! \frac{1}{N_t} \sum_{m \in \mathcal{A}_{t}}\! \tau_m \) is the average number of epochs. 

Let $\boldsymbol{\varepsilon}_{t}=\hat{\mathbf{s}}_{t} - \mathbf{s}_{t}$ represent the aggregation error as given in (\ref{error_term_exp}). Thus, using Assumption 4 and considering the zero-mean aggregation error, it can be concluded that
\begin{align}
 \mathbb{E}[\|\hat{\mathbf{s}}_{t}\|^2] \!=\! \mathbb{E}[\|\mathbf{s}_{t}\|^2] \!+\! \mathbb{E}[\|\boldsymbol{\varepsilon}_{t}\|^2] \!\leq\! \mathbb{E}[\|\mathbf{s}_{t}\|^2] + \zeta_t^2. \label{expshat2}  
\end{align}

Using (\ref{newideal}) and \cite[Assumption 1]{10109680}, we have \(\mathbb{E}[\langle \nabla F(\mathbf{w}_t), \hat{\mathbf{s}}_{t} \rangle]\!=\!\mathbb{E}[\langle \nabla\! F(\mathbf{w}_t), \mathbf{s}_{t}\!+\! \boldsymbol{\varepsilon}_{t}\rangle]\!=\!\eta \bar{\tau}_{t} \mathbb{E}[\|\nabla F(\mathbf{w}_t)\|^2]\). Thus, by substituting this, (\ref{exps2}) and (\ref{expshat2}) into (\ref{expmain}), we obtain
\begin{align}
&\mathbb{E}[F(\mathbf{w}_{t+1})]\nonumber\\ 
&\leq\mathbb{E}[F(\mathbf{w}_t)] - \eta \bar{\tau}_{t} \mathbb{E}[\|\nabla\! F(\mathbf{w}_t)\|^2] 
+ \frac{L}{2} \left[ \eta^2 \bar{\tau}_{t}^2 G^2 + \zeta_t^2 \right].\label{finalone}
\end{align}

In order to evaluate the convergence after $T$ communication rounds, using (\ref{finalone}) and the telescoping sum \cite{10109680,10696917}, it can be written that
\begin{align}
&\mathbb{E}[F(\mathbf{w}_{T+1})] - \mathbb{E}[F(\mathbf{w}_1)]\nonumber\\
&\leq - \sum_{t=1}^{T} \eta \bar{\tau}_{t} \mathbb{E}\left[\|\nabla F(\mathbf{w}_t)\|^2\right] 
+ \sum_{t=1}^{T} \frac{L}{2} \left( \eta^2 \bar{\tau}_{t}^2 G^2 + \zeta_t^2 \right).\!\label{avbound1}
\end{align}

To get the average convergence bound, we divide (\ref{avbound1}) by $T$ and rearrange the terms, which yields 
\begin{align}
&\frac{1}{T}\sum_{t=1}^{T} \eta \bar{\tau}_{t} \mathbb{E}\left[\|\nabla F(\mathbf{w}_t)\|^2\right]\nonumber\\
&\leq \frac{\mathbb{E}[F(\mathbf{w}_1)] - F^*}{T} + \frac{1}{T}\sum_{t=1}^{T} \frac{L}{2} \left( \eta^2 \bar{\tau}_{t}^2 G^2 + \zeta_t^2 \right),\label{bounT}
\end{align}
where $F^*$ is the optimum global loss value.

Since $\bar{\tau}_{t}$ varies across communication rounds due to energy constraints, we use its bounds as \(\hat{\tau}_{\min} \leq \bar{\tau}_{t} \leq \hat{\tau}_{\max}\) to account for epoch variability in the convergence analysis, where summing over $T$ rounds gives \(T\hat{\tau}_{\min} \leq \sum_{t=1}^{T}\bar{\tau}_{t} \leq T\hat{\tau}_{\max}\). Moreover, we define \( \zeta^2 = \max_t \zeta_t^2 \) to avoid tracking per-round aggregation noise and simplify the analysis. This bounds the average as \(\frac{1}{T} \sum_{t=1}^{T} \zeta_t^2 \leq \zeta^2\). Therefore, by applying these bounds to (\ref{bounT}), the convergence bound is obtained as
\begin{align}
\frac{1}{T} \sum_{t=1}^{T} \mathbb{E}\left[\|\nabla F(\mathbf{w}_t)\|^2\right]
\leq \frac{\Delta_0}{\eta T \hat{\tau}_{\min}} 
+ \frac{L \eta \hat{\tau}_{\max} G^2}{2}
+ \frac{L \zeta^2}{2 \eta \hat{\tau}_{\min}},\label{finalconv}
\end{align}
where $\Delta_0 \triangleq\mathbb{E}[F(\mathbf{w}_1)] - F^*$.
\begin{remark}
From (\ref{finalconv}), it is evident that faster convergence is achieved by improving the number of epochs and aggregation error. In the proposed framework, the number of epochs is improved via Algorithm \ref{adalg}, which dynamically allocates the epochs with full or fractional datasets based on the available energy. Simultaneously, \(\zeta^2\) is reduced through the proposed denoising strategies that mitigate the effects of fading, CCI, and AWGN. Moreover, the first term in the bound decreases with the \(T\), leading to a convergence rate of \(\mathcal{O}(1/T)\). 
\end{remark}
\section{Simulation Results}\label{sec5}\vspace{-5pt}
Following the setup in \cite{9264742}, we consider a \(200 \times 200\) m square area with the PS at the center \((0, 0)\), where devices are uniformly distributed with coordinates \((x_m, y_m) \in [-100, -20] \cup [20, 100]\) and distances \(d_m = \sqrt{x_m^2 + y_m^2}\), inband nodes with \((x_i^{\mathrm{in}}, y_i^{\mathrm{in}}) \in [-140, -120] \cup [120, 140]\), \(d_i^{\mathrm{in}} = \sqrt{(x_i^{\mathrm{in}})^2 + (y_i^{\mathrm{in}})^2}\), and \(d_{m,i}^{\mathrm{in}} = \sqrt{(x_i^{\mathrm{in}} - x_m)^2 + (y_i^{\mathrm{in}} - y_m)^2}\), and outband nodes with \((x_k^{\mathrm{out}}, y_k^{\mathrm{out}}) \in [-100, -25] \cup [25, 100]\) and \(d_{m,k}^{\mathrm{out}} = \sqrt{(x_k^{\mathrm{out}} - x_m)^2 + (y_k^{\mathrm{out}} - y_m)^2}\). We set the required parameters as \(M=\{10,25,50,100\}\), $\delta_m=0.9$, $\xi=2.5$, \(I=K=100\), \(P^{\mathrm{in}}_i=P^{\mathrm{out}}_k=0.1\) W, \(P_{m,t}^{\text{up}}=10\) dBm, \(T^{\text{h}}=1\) sec, \(N_{0}=-80\) dBm, \(B_{\max}=B_{m,1}=\) 50 J\cite{9453711}, \(\eta=0.01\), \(d=582026\), \(\tau=2\) for non-adaptive local training cases, \(\kappa=10^{-28}\), \(C_m=1.3\times 10^4\) cycles/sample, and \(f_m= 2 \) GHz \cite{9264742}, unless otherwise stated. Moreover, we evaluate the performance of the proposed FL system on the MNIST image classification task, where the dataset of handwritten digits (0–9) is independently and identically distributed across devices. Each device is allocated 1,200 training samples and performs local updates using a convolutional neural network with the same model architecture as that used in \cite{9672092}, for one or more epochs per round, depending on its available energy.

\begin{figure}[t]
\centering\includegraphics[width=0.8\linewidth]{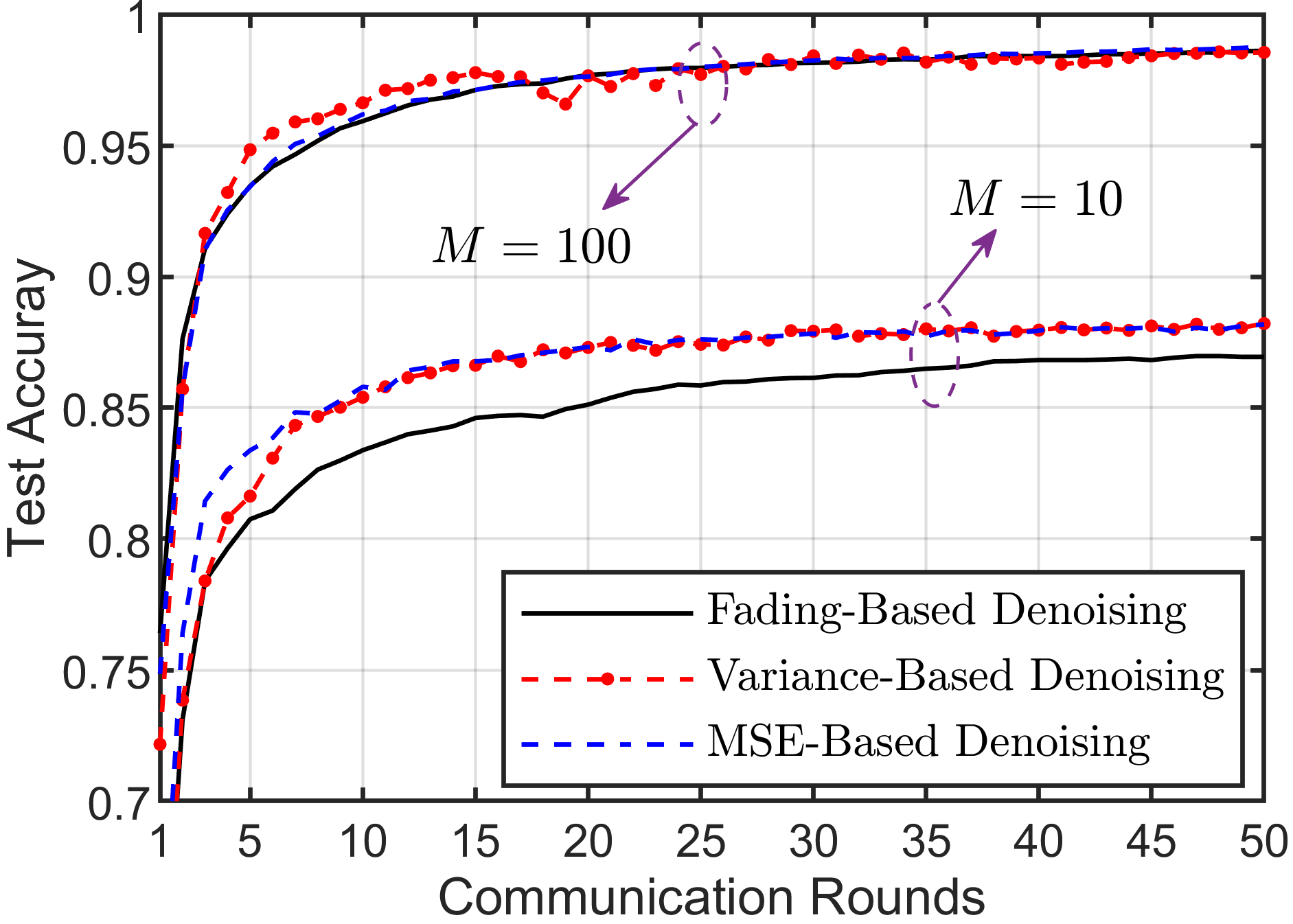}\par \caption{\ Test accuracy under different denoising policies.}\label{Fig1}\vspace{-20pt}
\end{figure}

Fig. \ref{Fig1} presents a comparison of the test accuracy achieved using the proposed variance-based denoising against the baseline fading-based and MSE-based denoising policies. As expected, the test accuracy improves with increasing the communication rounds for all methods. Notably, for a small number of devices (e.g., $M=10$), the variance-based approach outperforms the fading-based denoising. This is because, for small $M$, the summation term in (\ref{optden}) becomes insufficient to effectively mitigate the CCI and AWGN. Furthermore, regardless of the number of devices, the variance-based scheme achieves accuracy comparable to the MSE-based approach while eliminating the need for CSI, demonstrating its effectiveness and scalability for large-scale OTA FL systems.

To validate the robustness of the proposed adaptive algorithm (Algorithm~\ref{adalg}), we compare its performance against conventional methods, namely non-adaptive schemes with and without energy storage, each operating with a fixed number of epochs. In the non-adaptive without energy storage method, a device becomes active and performs a fixed number of epochs only if it satisfies the energy constraint; otherwise, it remains idle and the harvested energy is discarded. In contrast, the non-adaptive with energy storage method allows devices to store unused energy in a battery when they are unable to participate due to insufficient energy, enabling them to accumulate energy for use in future communication rounds. 

\begin{figure}[t!]
  \centering
  \begin{subfigure}[b]{\linewidth}
    \centering
    \includegraphics[width=0.8\linewidth]{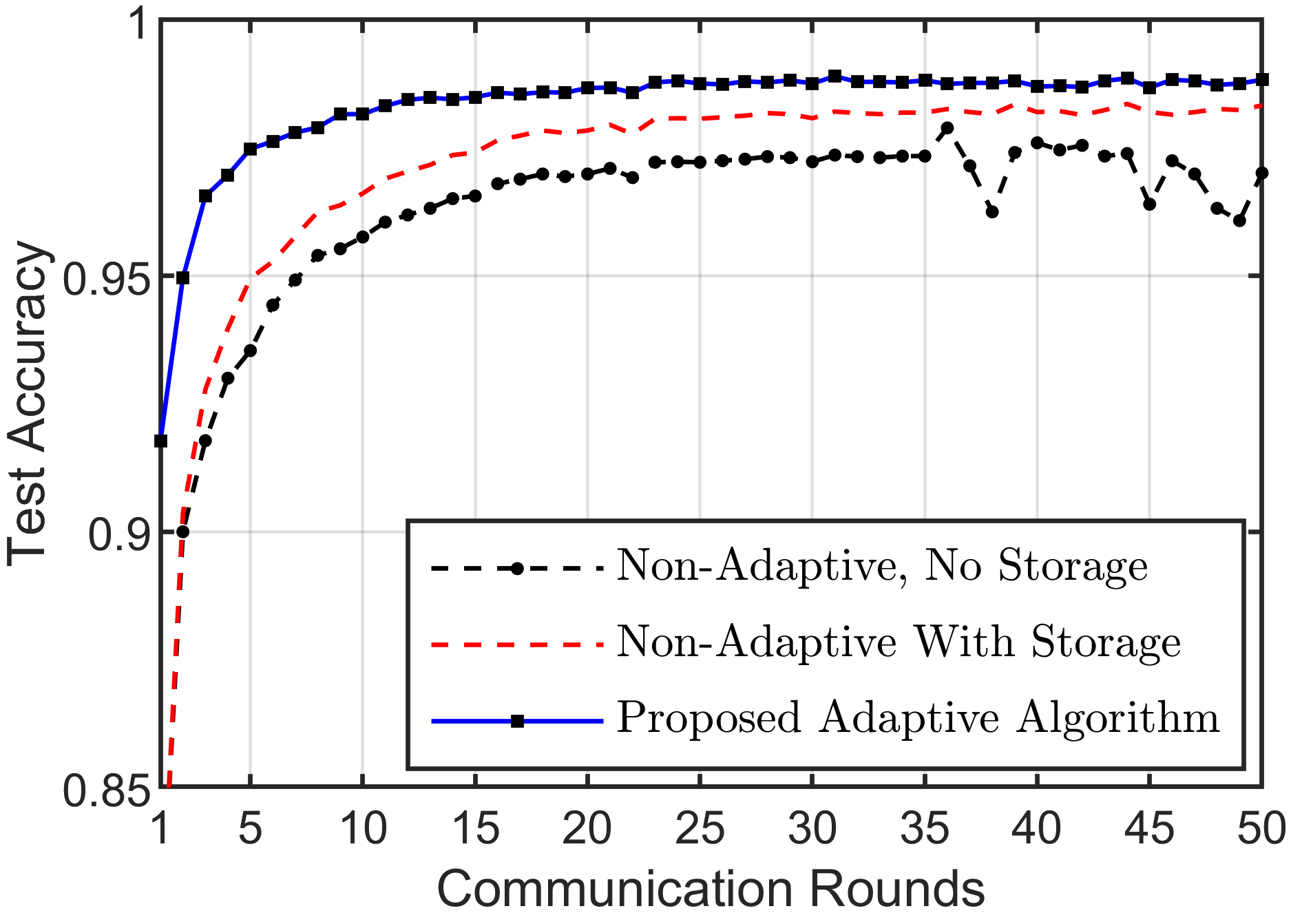}
   \vspace{-5pt}\caption{}
    \label{Fig2a}
  \end{subfigure}\vspace{5pt}
  \begin{subfigure}[b]{\linewidth}
    \centering
    \includegraphics[width=0.8\linewidth]{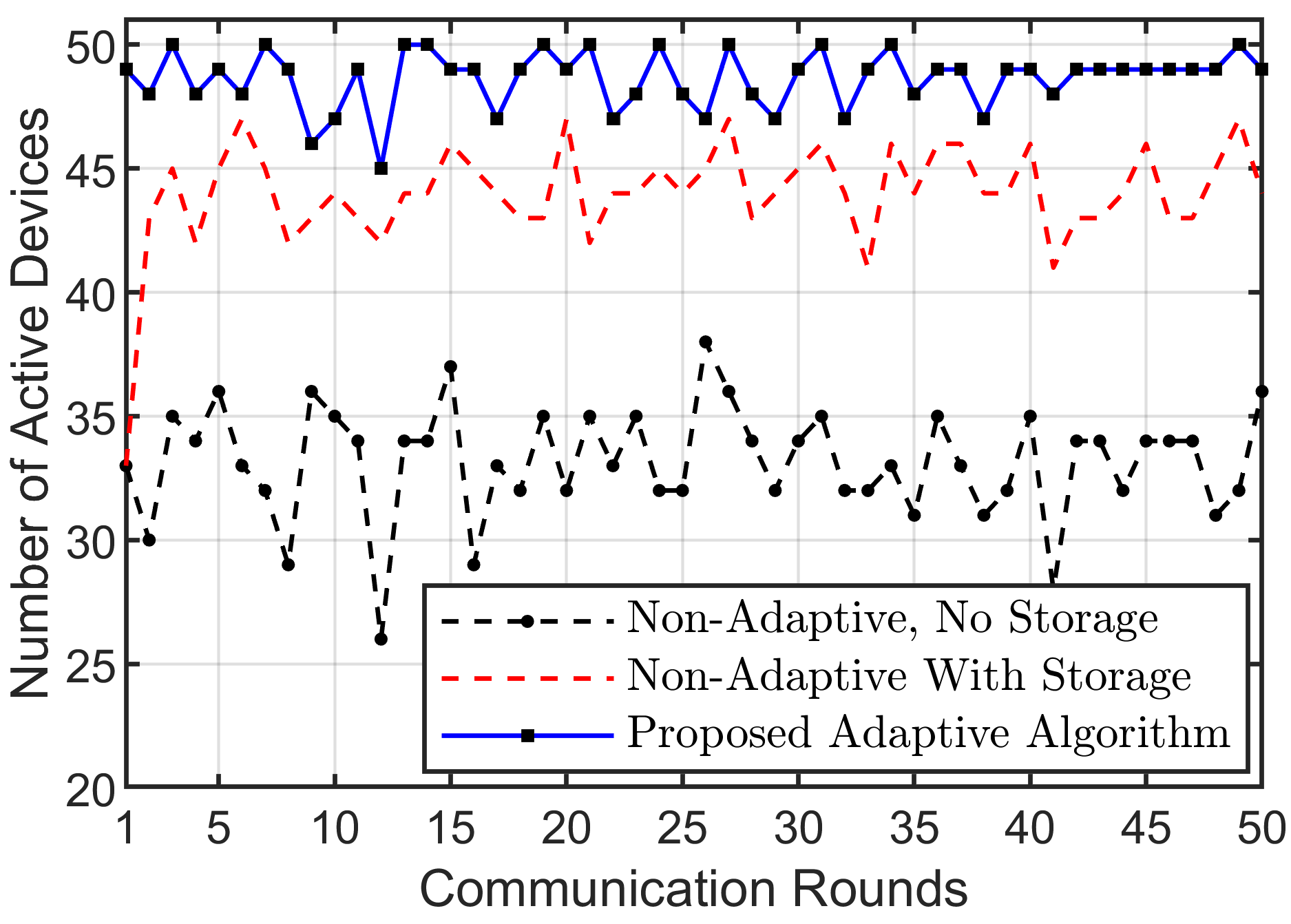}
     \vspace{-5pt}\caption{}
    \label{Fig2b}
  \end{subfigure}\vspace{5pt}
  \begin{subfigure}[b]{\linewidth}
    \centering
    \includegraphics[width=0.8\linewidth]{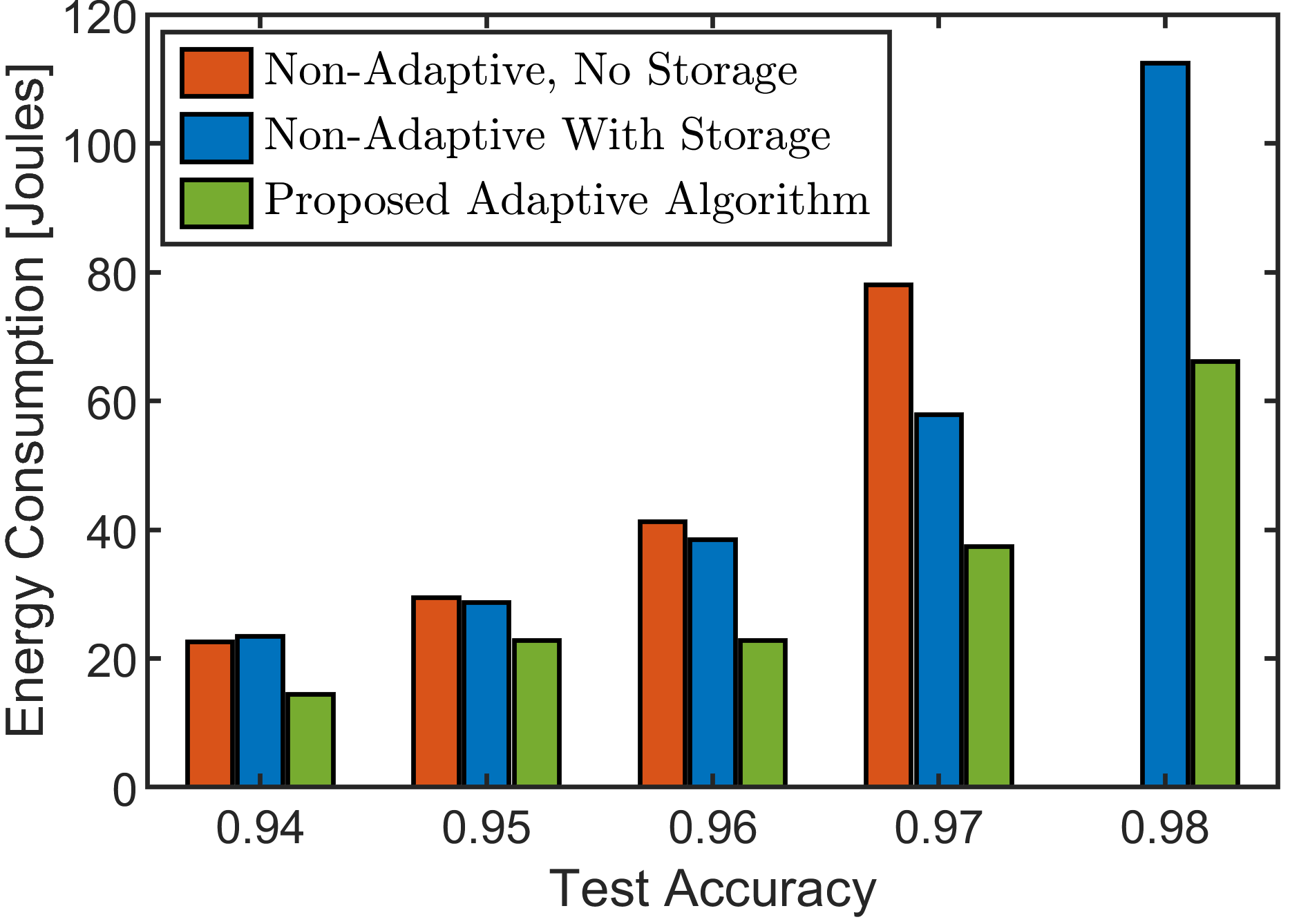}
    \vspace{-5pt}\caption{}
    \label{Fig2c}
  \end{subfigure}
  \vspace{-15pt}
  \caption{\ Effects of adaptive algorithm on the OTA FL performance: (a) Test accuracy, (b) Device participation, and (c) Energy consumption.}
  \label{Fig2}\vspace{-10pt}
\end{figure}

Fig. \ref{Fig2} (a) compares the test accuracy performance of the proposed adaptive algorithm with the two non-adaptive baselines under the same EH conditions. As observed, the adaptive method consistently outperforms both non-adaptive approaches across all communication rounds. This improvement is attributed to the algorithm's ability to dynamically adjust the number of epochs and the fraction of the dataset used based on each device’s available energy. Therefore, such flexibility enables faster convergence and higher final accuracy. 

Fig. \ref{Fig2} (b) illustrates the number of active devices per communication round under the proposed adaptive algorithm and the two non-adaptive baselines. The adaptive method consistently achieves higher device participation across rounds. This is primarily due to its flexible scheduling mechanism, which allows devices to contribute updates even with limited energy. Unlike the non-adaptive schemes that require the devices to run a fixed number of epochs, causing the devices to remain idle if they cannot meet the energy demand, the adaptive algorithm checks whether a device can perform at least one epoch. If full training is still infeasible, it further enables participation using a reduced subset of the local dataset with one epoch. This dual-level adaptation—adjusting both the number of epochs and the data size—significantly increases the number of active devices. As a result, more devices are able to contribute updates in each round, which directly supports the improved accuracy trends observed in Fig. \ref{Fig2} (a). 

Fig. \ref{Fig2} (c) compares the total energy required to reach specific accuracy levels for the proposed adaptive algorithm and two non-adaptive baselines. The adaptive method is clearly more energy-efficient across all accuracy targets. This efficiency is due to two main reasons. First, by adjusting the number of epochs and allowing partial dataset training, the adaptive algorithm enables devices to contribute updates with minimal energy. Second, and more importantly, the adaptive strategy converges significantly faster, requiring fewer communication rounds to reach a given accuracy. Since each communication round incurs uplink transmission energy, this leads to substantial savings in communication energy. In contrast, the non-adaptive scheme with energy storage retains energy for future computation, avoiding wastage, but requires more communication rounds due to inflexible scheduling—thus incurring a higher total transmission energy. The non-adaptive scheme without storage performs worst due to both energy waste and slower convergence. These results confirm that Algorithm~\ref{adalg} achieves better accuracy with less total energy consumption, making it well-suited for energy-constrained FL systems.

\begin{figure}[t]
\centering\includegraphics[width=0.8\linewidth]{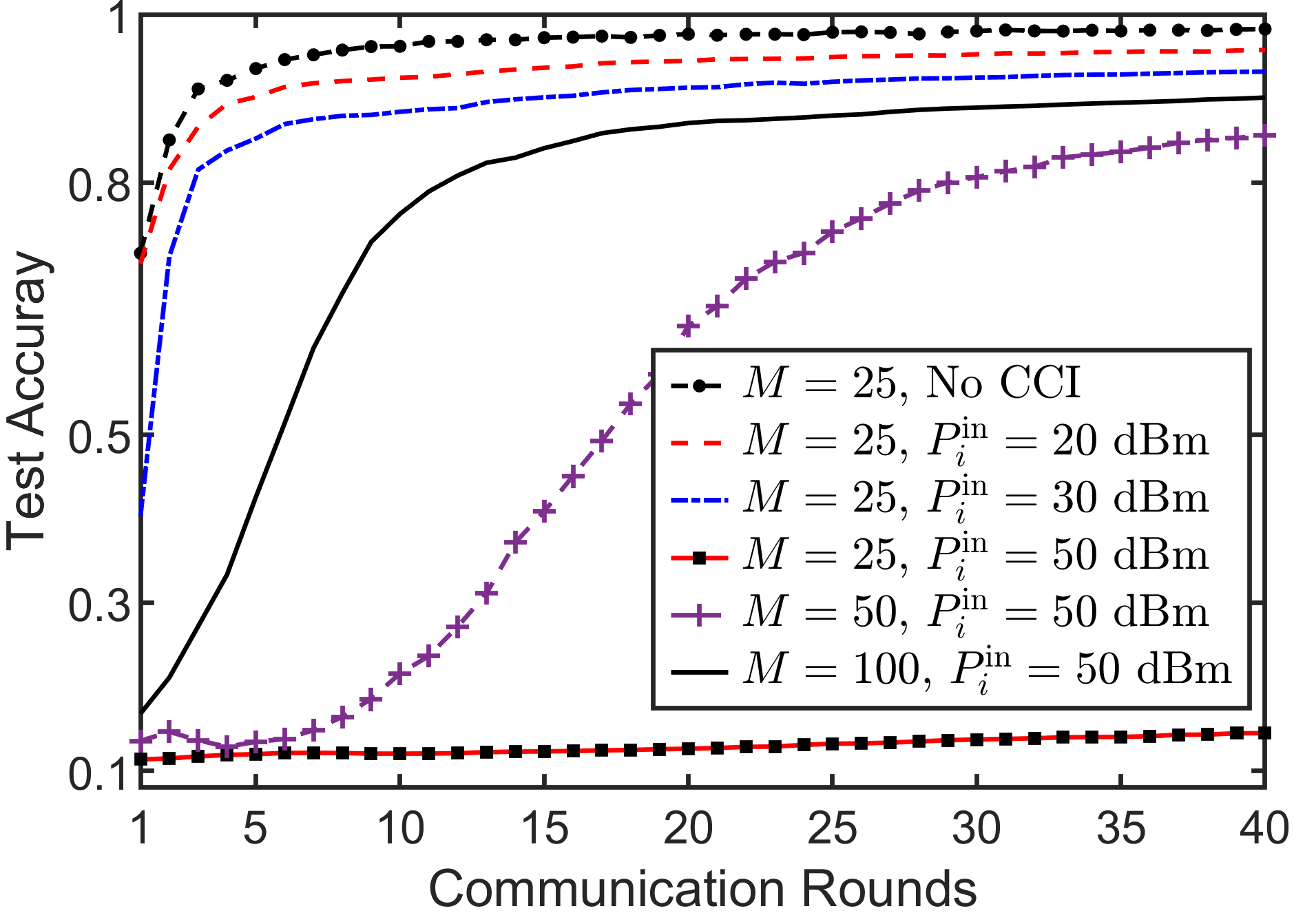}\par \caption{\ Effects of CCI on the test accuracy under different $M$.}\label{Fig3}\vspace{-15pt}
\end{figure}

Fig. \ref{Fig3} illustrates the impact of CCI and the number of participating devices on the test accuracy. In the absence of CCI, the system achieves near-optimal accuracy, serving as a performance upper bound. However, as the CCI power increases, particularly at \(P^{\mathrm{in}}_i = 50\) dBm, the performance significantly deteriorates due to increased OTA aggregation error. Notably, increasing $M$ from 25 to 50 or 100 enhances resilience to interference, as more device updates help average out the noise and interference. This observation underscores the importance of device participation in mitigating the impact of CCI. Since the proposed adaptive algorithm increases the number of active devices per round, it can indirectly improve robustness against interference.
\section{Conclusion}\label{sec6}
This paper investigated analog OTA FL with EH-based devices in the presence of CCI. To address practical system limitations, we proposed a CSI-free variance-based denoising policy and an adaptive scheduling algorithm that dynamically adjusts the number of epochs and dataset size based on the available energy. Simulation results demonstrated that the proposed denoising policy performs comparably to CSI-based methods while avoiding their complexity. Moreover, the adaptive algorithm significantly improves device participation, accelerates convergence, and reduces the total energy consumption. Notably, it also enhances robustness against CCI by enabling more devices to contribute updates. These results highlight the effectiveness and scalability of the proposed techniques for energy-constrained OTA FL systems.
\bibliographystyle{IEEEtran}
\bibliography{IEEEabrv,FLREF}

\end{document}